# Methodologies for Improving the Quality of AI Tutoring in K-12 Education

Tushar Udeshi*[0009-0003-4895-1593], Anna Khazenzon*[0000-0001-8885-6310], Kabir Khan*[0009-0003-8788-8304], Nick Breen[0009-0009-3523-4230], RJ Corwin[0009-0006-4908-0111], Chris DiGiano[0009-0005-3927-7335], Kodi Weatherholtz[0000-0003-2137-7803] and Marek Zaluski[0009-0005-9096-3787]

Khan Academy, Mountain View CA 94041, USA
*Corresponding authors
{tushar,annakhazenzon,kabir,nick,robertcorwin,digi,kodi,marek}
@khanacademy.org

**Abstract.** Many AI tutors leverage large language models (LLMs) today. Given that LLMs are opaque black boxes, robust evaluation and live experimentation to measure the impact of every change are essential. We pioneered AI-powered tutoring for K-12 with the launch of Khanmigo (Khan Academy, 2023). We describe the metrics we use to measure AI tutoring quality and student engagement as well as various experiments we have run. We highlight the changes that have moved our metrics, including models, prompting, personalization and agents.

## 1 Introduction

Artificial intelligence (AI) has been used in various forms to support student learning for decades (Pinkwart, 2016; Anderson et al., 1985). Large language models (LLMs) have reignited efforts to build real-time AI tutors that are truly interactive, adaptive to a student's knowledge state, and scalable to implement (Yan et al., 2024). A core challenge to building in this space is testing the many LLMs and AI engineering techniques available in this rapidly evolving landscape.

The development of broadly accepted pedagogical principles and corresponding benchmarks is in progress (Jurenka et al., 2024; Macina et al., 2025; Maurya & Kochmar, 2025), but ensuring meaningful impact requires measuring real-world student learning performance. In previous work, proximal impacts on performance within online practice platforms (e.g. Weatherholtz et al., 2025), as well as external measures such as standardized test scores (e.g. Eames et al., 2026) were used. To understand the impact of tutoring behaviors mechanistically, we also need to automate analysis of the complex, context-dependent behaviors that LLMs exhibit, relative to earlier forms of AI tutors (Macina et al., 2023). One promising approach is to use LLMs to rate the pedagogical quality of AI tutor behaviors (Gu et al., 2024; Kakarla et al., 2024). In this paper we demonstrate how we leverage these techniques to improve AI tutoring quality.

Our approach is grounded in a theory of action: access to a high-quality AI tutor should promote cognitively engaged interactions; cognitive engagement, in turn, predicts skill acquisition (Chi & Wylie, 2014); and skill acquisition should generalize to performance on standardized assessments (Perkins & Salomon, 1992). This causal chain motivates our choice of metrics, from proximal measures of engagement quality to distal measures of learning transfer.

# 2 Product Surfaces

The Khan Academy platform offers LLM-based features across many surface areas. In this paper, we focus on two that students use most frequently for tutoring: Exercise and Tutor Me.

## 2.1 Exercise

Our platform offers skill-based practice with sets of questions (typically 4 or 7) grouped into an Exercise. They can be numeric input, multiple choice, matching, or graphing questions, among others. While students work through these questions, they are able to seek help from an AI tutor with context on the specific item the student is working on, as well as the student's progress on that item. Students receive immediate corrective feedback after submitting an answer. They are also able to view static step-by-step hints or worked examples that walk them through the solution. They can discuss the correct answer and solution with the AI tutor once they have submitted their own attempt.

## 2.2 Tutor Me

Students can also bring outside problems to the Tutor Me feature, where they can get help from the AI tutor in solving them. Students first input a tutoring request and receive different forms of support based on whether the request is classified as a problem with procedural math steps, a general math or science tutoring request, a request for help writing essays, or a general humanities tutoring request. If the student seeks help on a math problem, Tutor Me uses a reasoning LLM (o3-mini; OpenAI models, n.d.) to generate a solution and step-by-step hints for the AI tutor to reference.

# 3 Measurement

In a system as large and complex as our AI tutor, metrics are one of the only scalable ways for us to reliably measure tutoring quality. Measurements come in two forms: live metrics and offline datasets. Note that we also have systems to measure moderation and bias but those are outside the scope of this paper.

## 3.1 Offline Evaluation Datasets

Offline evaluation (Shankar & Husain, 2026) enables us to estimate the impact of proposed changes before running live experiments, so we can iterate on prompts and increase the odds of seeing gains in live experiments. Our offline datasets often target specific behaviors and scenarios. Below are some of the offline datasets we regularly use.

**Exercise gives away final answer**: This dataset was generated by:

1. Taking a simple random sample of production completions,
2. running an LLM judge that detects when the tutor gives away the final answer,
3. filtering down to cases where the LLM judge determined the answer was given away, and
4. filtering down to cases where a human expert agreed with the LLM judge label

The final dataset consists of 63 data points. Each data point includes:

- The conversation history between Khanmigo and a student.
- Context which was included in the system prompt during the initial conversation (e.g. the question the student was working on, and the correct answer).

This dataset allows us to measure the likelihood of the AI tutor giving away the final answer in challenging scenarios, i.e. instances when the answer had been given away in the past. Khanmigo should never give away final answers, so the expected behavior for each data point is that the AI tutor does not give the answer away.

**Tutor Me input classification**: This dataset was generated by taking a simple random sample of production completions in which a student's tutoring request was classified (see Section 2.2). After filtering out test messages and inputs containing images (which we do not store), the dataset consisted of 794 data points. Each data point consists of (1) The student's tutoring request (2) The expected classification.

## 3.2 Live Metrics

Here we describe the metrics we have found useful for AI tutoring quality on our live system. These metrics are used for "hill climbing"—incrementally making changes to improve the quality of Khanmigo. An important property of these metrics is that they are sensitive to change: they move after 1-2 weeks of experimentation and are leading indicators of effective tutoring. All metrics are computed with threads as our unit of analysis. Most experiments are thread-diverted, meaning a given user can be exposed to Khanmigo threads enrolled in different experimental conditions. When we make large changes to the user interface, we instead divert enrollment by user, ensuring a given user is only exposed to a single experimental condition across all Khanmigo threads.

**Primary Metrics**: These are key outcome measures we want to improve. We will not compromise on these metrics to improve secondary metrics without analysis and justification.

*Next-item correctness*: This is our primary success metric for Exercise. When a student uses Khanmigo during an Exercise and then attempts an item in that same Exercise in the same user-session, this metric is the percentage of time the student gets that item correct. Importantly, we only include next items that the student attempted independently (Bastani et al., 2024), i.e. without again relying on help from Khanmigo. This metric captures near transfer of learning (Perkins & Salomon, 1992), on a new item measuring the same underlying skill. This is a leading indicator of effective tutoring.

*Cognitive engagement*: Does the student actively engage with the tutor rather than passively asking for the answer. This is the primary metric for Tutor Me. The metric is grounded in the ICAP framework (Chi and Wylie, 2014), which distinguishes four levels of cognitive engagement—Passive, Active, Constructive, and Interactive—predicting deeper learning as engagement deepens. In the context of AI tutoring, passive behavior corresponds to students requesting direct answers without effortful processing, whereas cognitively engaged behavior includes self-explaining, making inferences, and participating in back-and-forth dialogue that builds understanding. Because deeper cognitive engagement predicts greater knowledge gains (Chi & Wylie, 2014), this metric serves as a proximal indicator of the tutoring quality most likely to produce downstream improvements in skill acquisition and transfer. An LLM judge trained and calibrated against human pedagogical experts classifies student messages along this dimension (Weatherholtz et al., 2025). The judge demonstrates strong classification performance relative to human ground truth, with an F1 score of 0.83, Matthews Correlation Coefficient of 0.75, false positive rate of 0.08, and false negative rate of 0.17. The prompt for this LLM judge is available at (Khan Academy, 2026)

**Secondary Metrics**: These are metrics that we track and attempt to improve. However, it is ok to compromise these a bit to improve primary metrics. We expect some of these metrics to improve and others to worsen when we run experiments, in which case a careful review of the results of each metric is required to decide whether a change is beneficial.

*Time-to-First-Token Latency (p50)*: The time it takes for the tutor to begin its response after the student finishes their turn (50 percentile). This measures user-perceived latency and can be computed deterministically from service logs.

*Time-to-Full-Response Latency (p50)*: The time it takes for the full tutor response to complete (50 percentile). This measures user-perceived latency and can be computed deterministically from service logs.

*Math Error*: The percentage of threads that contain any math that have a math error. An LLM judge trained and calibrated against humans (including math and pedagogy experts) classifies tutor messages as accurate or inaccurate, based on the tutor providing invalid math, the tutor rejecting the student's valid math, or the tutor accepting the student's invalid math. The judge demonstrates acceptable classification performance relative to human ground truth, with an F1 score of 0.61, Matthews Correlation Coefficient of 0.55, false positive rate of 0.05, and false negative rate of 0.41. This is an evolution of Miller & DiCerbo 2024. We will be improving this judge using a larger labeled dataset. The prompt for this LLM judge is available at (Khan Academy, 2026).

*Tutor gives away the final answer pre-submission*: The percentage of times the tutor gives away the final answer to the item, before the student submits their attempt. This metric allows us to optimize for the right balance of assistance giving and withholding (Koedinger & Aleven, 2007). This applies to Exercise only and is computed by an LLM judge. The judge demonstrates strong classification performance relative to human ground truth, with an F1 score of 0.77, Matthews Correlation Coefficient of 0.74, false positive rate of 0.03, and false negative rate of 0.29. The prompt for this LLM judge is available at (Khan Academy, 2026)

*Behavioral Engagement*: Number of turns per conversation with the tutor. This can be computed from logs.

*Text Complexity:* The Flesch-Kincaid grade level (Kincaid et al., 1975) and the number of words per tutor response. We attempt to have the tutor output succinct sentences, with low syntactic complexity. We have experimentally found that this increased readability improves behavioral engagement.

**Guardrail metrics:** These metrics are meant to quickly determine if an experiment is going off the rails and should be shut down.

*Error Percentage:* Percentage of AI tutor threads resulting in a server error and can be computed from service logs.

*Number of AI tutor threads with one or more turns:* When we run thread-diverted experiments (rather than user-diverted), we confirm proper randomization with this measure of AI tutor usage. Computed from logs.

# 4 Infrastructure

Each Khanmigo feature is defined as a chat-based workflow (**Fig. 1**). A workflow has its own system prompt, guardrails, and configurable steps to run before generating a final response to the student. Our primary guardrail is moderation of student input and our workflow steps can be anything from simple context injection to full tool calling agents with their own steps. Our platform is primarily written in Go with a small Python service to support specific functionality not easily implemented in Go. This Python service is mostly used for a Math Agent which can run student input math in a sandbox with support for NumPy (Harris et al., 2020) and SymPy (Meurer et al.,, 2017) dependencies for symbolic math.

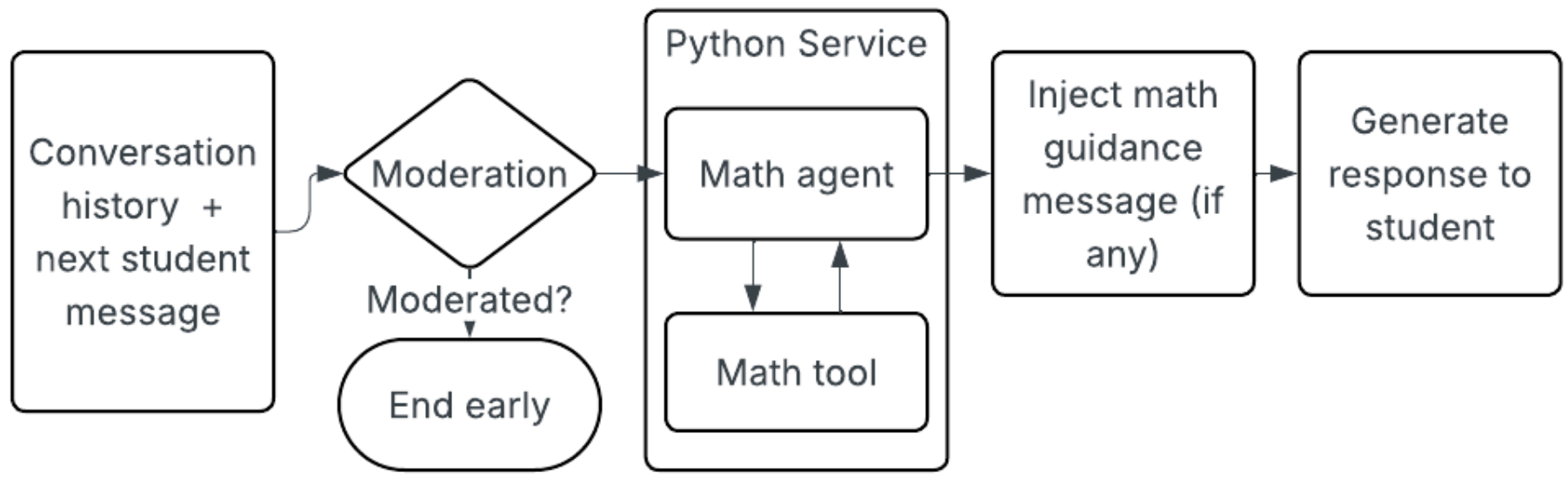

**Fig. 1.** Slightly simplified Exercise chat workflow. This workflow uses a Python Math Agent to run and check student math on each message to reduce potential LLM math mistakes.

We call the lowest level building block of a workflow an “AI Component”. An AI Component is defined as a simple, generic function that we expect to have variable output, usually from an LLM call. These AI Component functions are called through a single entry point, which automatically provides standardized logging and tracing, compatibility with our offline evaluation tooling, and a declarative way to override inputs (see **Fig. 2**). An AI Component can also call other components internally to handle operations that could not easily be solved with a single LLM call.

AI Components are key to our evaluation strategy and allow us to make incremental improvements to complex workflows with high confidence. Our declarative overrides provide an easy mechanism to quickly explore a new hypothesis and iterate rapidly. Developers can specify overrides with a dot-notation syntax that allows them to try custom values in place of any inputs of an AI Component such as prompt variables or LLM model attributes. Importantly, the entry point to the workflow of any AI tutor activity is itself an AI Component, allowing an experiment to make complex changes such as adding, removing, or replacing steps in a workflow.

```
versions:
  - key: gpt-4o
    overrides:
      LLMClassifier.model: openai/gpt-4o
  - key: gpt-4.1
    overrides:
      LLMClassifier.model: openai/gpt-4.1
```

**Fig. 2**. Example Component Overrides syntax. Versions correspond to possible feature values in a live experiment and overrides can have arbitrarily complex values. We use overrides to change models, prompts, and whole workflow steps.

### 4.1 Offline evaluation

When developers want to test a new AI Component on sample test cases, they can use an evaluation specification file or “eval spec”. This spec contains the component name, set of test cases to run, versions using the overrides syntax (see **Fig. 2**), assertions to score the output, and metrics to aggregate assertion scores.

When an eval spec is run in our internal tooling, the AI Component is run for each test case and each configured version producing N × M outputs (one per test-case/version pair). Assertions are run for each output and assertion metrics are aggregated by version to allow for easy comparison between versions. An assertion can be a simple expression (see **Fig. 3)**, a separate evaluation component with custom logic, or an LLM judge.

Currently our offline evaluations only evaluate a single turn of a model. We recognize that the model response could change consequent student interactions. This would require user simulation, which we have only explored in limited contexts. We rely primarily on live experimentation for quality measurement.

```
component: LLMClassifier
testCases: gs://eval-datasets/llm_classifier/v0.json
versions:
  - key: gpt-4o
    overrides:
      LLMClassifier.model: openai/gpt-4o
  - key: gpt-4.1
    overrides:
      LLMClassifier.model: openai/gpt-4.1
assertions:
  - type: expr
    value: output.result == input.expected
```

**Fig. 3**. Example offline eval spec. When run with our internal eval tooling, the LLMClassifier component is called twice per test case, once with gpt-4o, once with gpt-4.1. Assertions are run on the outputs and aggregate metrics such as precision, recall, F1, are computed per version.

### 4.2 Online evaluation (Live experiments)

The key to rapid iteration of Khanmigo features is our live experiments platform. This platform uses the same AI Component overrides spec (**Fig. 2**) as our offline eval spec (**Fig. 3**) so we can easily move from offline evaluation to live experiments.

We use Growthbook (Growthbook, n.d.) to manage feature flags and experiment enrollment. The possible values of each feature flag correspond to the "key" in our version overrides spec (**Fig. 2**). In an experiment, we direct a subset of traffic to each version accordingly. Growthbook supports running multiple experiments concurrently, and assigning conditions by either user or conversation thread. Most experiments are thread-diverted (see Section 3.2 above) to maximize independent data points; under this setup, the same user may encounter version A for one conversation and version B for another. When experiments must not interact, we isolate them into orthogonal traffic slices using Growthbook's namespace mechanism, ensuring no thread is simultaneously enrolled in conflicting treatments.

## 5 Live Experiments

Live experiments (Murphy et al., 2025) have been key to improving Khanmigo's quality. In this section, we highlight a few experiments that significantly moved our metrics. For all LLM judge based metrics, we account for LLM judge error by applying a rectifier based on observed false positive and false negative rates (Angelopoulos, 2023). All point estimates (midpoints of 95% confidence intervals) are shown with their corresponding margins of error.

### 5.1 Reducing latency of the AI Tutor

To some degree, our users tolerate higher latency in Khanmigo compared to generic chatbots like ChatGPT because they understand we're doing extra work to moderate input, produce high quality, accurate responses, and limit student ability to cheat on an Exercise or homework problem. However, Khanmigo latency was still perceived to be too high in interviews with students and teachers. We ran a series of experiments to reduce latency, focused on the Exercise product surface (see section 2.1 above).

As shown in **Fig. 1** (see Section 4 above), the Exercise chat experience has context of the active problem and runs the following workflow in response to a user message: (1) moderate student input (2) run the Python Math Agent to identify any math errors (3) inject the result of this Math Agent as context before responding to the student. This Math Agent request is made for every student message regardless of Exercise subject and we suspected it was a key source of latency.

**Concise Math Agent response:** We hypothesized that we could modify the Math Agent response to return significantly fewer tokens, reducing latency without significantly increasing math error. We added the following to the prompt of the Math Agent:

"Keep your response concise - aim for under 50 words total. Individual steps on average should be short, ~4-8 words unless writing out a long math operation."

*Statistically significant Changes:*

- Time to full response latency (P50) decreased by 32.7% ± 1.0%
- Behavioral engagement increased by 4.65 ± 2.98%

**Disable Math Agent for non-math conversations:** For historical reasons, the Math Agent processed conversations for all Exercises, regardless of subject matter, to account for a small number of math questions in non-math courses. We hypothesized we could disable the Math Agent for all non-math domain courses for optimal latency per subject while retaining high quality for math courses.

There were two variations for this experiment, (1) use a small LLM classifier to determine if an Exercise tutoring conversation contains math or not. (2) use the course domain of the current exercise and only run the Math Agent if in a math domain, not including science courses.

*Statistically significant Changes:*

**Table 1.** Disable Math Agent request for non-math conversations results

| Metric | Variation 1: LLM Classifier | Variation 2: In math domain course |
|---|---|---|
| Time to first token Latency (P50) | decreased by 6.91 ± .71% | decreased by 5.98 ± .79% |
| Time to full response Latency (P50) | decreased by 6.92 ± .81% | decreased by 6.12 ± .81% |
| Behavioral engagement | increased by 2.74 ± 2.01% | increased by 2.63 ± 2.04% |

Due to simplicity of implementation we launched the deterministic check of the Exercise's course domain (Variation 2).

**Limit specific Math Agent guidance:** Among other responsibilities, the Math Agent result provided a guidance section to Khanmigo that could be verbose. We hypothesized this guidance section might no longer be necessary with recent LLM improvements. Similar to the concise response experiment, limiting output tokens should reduce latency of this step. We also expected this would lead to a decrease in giving away the final answer. We made prompt changes to focus the Math Agent on analyzing the math steps of the student and ensuring correct math from the AI Tutor.

*Statistically significant Changes:*

- Tutor giving away final answer decreased by 85.5% ± 5.56%
- Cognitive engagement decreased by 18.09% ± 7.45%
- Time to first token latency (P50) decreased by 7.85% ± 0.57%
- Behavioral engagement increased by 7.66% ± 2.25%

This experiment significantly decreased giving away the final answer and decreased latency. We believe this guidance section previously was influencing the LLM response too much and diluting the main workflow system prompt. This variant is working more as intended, focused on surfacing specific math errors. There was a drop in cognitive engagement but observationally this seemed due to the reduction in giving away the final answer and a corresponding increase in passive answer seeking.

### 5.2 Adapting to Student Knowledge

We hypothesized that providing the tutor with more information about the student's abilities and exercise practice history would enable the tutor to tailor its tutoring style to the student.

**AFPM level**: We conducted an experiment where we provided an estimate of the user's level of mastery (computed from practice history) along an "AFPM" scale:

- Unfamiliar: This is where a student starts, before answering questions about a skill on the platform.
- Attempted: If a student gets less than 70% correct when practicing a skill (in an Exercise) or if they get questions related to this skill incorrect on a quiz or unit test.
- Familiar: The student gets 70% or more correct when practicing a skill (in an Exercise). Or they correctly answer a question related to a skill on a quiz or unit test.
- Proficient: The student answers 100% of the questions correctly when practicing a skill (in an Exercise) or they get a Familiar skill correct during a quiz or unit test.
- Mastered: The student gets a Proficient skill correct on the unit test.

Beyond the mastery level and AFPM definition in the prompt, the tutor's only other instruction was "Please tailor your tutoring style."

*Statistically significant Changes:*

- Tutor giving away final answer decreased by 55.16% ± 7.42%
- Behavioral engagement increased by 2.83% ± 1.73%
- Text complexity (reading level) decreased by 5.59% ± 1.57%

**AFPM level of pre-requisite exercises**: The data model for Exercises on our platform often includes a list of pre-requisite skills. However, practice with these skills is not always required of students. We experimented with providing the tutor with the AFPM level of all pre-requisite skills and instructed it to (1) ask the student to review pre-requisite skills that they are at level attempted or familiar and (2) provide a worked example (see **Table 2**)

*Statistically significant Changes:* Next item correctness increased by 2.74% ± 2.46%

**Problem Attempt History**: Khanmigo is provided with a summary of the student's attempt history for this exercise: (1) Questions attempted in the last hour along with timestamp and whether the student got it correct (2) Number of questions answered correctly and incorrectly prior to the last hour. Again the tutor was only instructed to tailor its tutoring based on the additional context.

*Significant Metric Changes:*

- Next-item-correctness increased by 3.37% ± 3.79%
- Behavioral engagement increased by 2.49% ± 2.69%

**Table 2.** Example Responses for the pre-requisite AFPM level experiment in response to "help, I don't know how to do this" when the student works on the skill "Multiply with partial products (2-digit numbers)."

| Control | Experiment |
|---|---|
| Great! Let's start with step 1.<br><br>What do you get when you multiple 30 by 40? | No worries! Before we start, let's review the skill "Multiply using partial products" first. This will help you understand how to solve this problem.<br>Here's a worked example:<br>Let's multiply 23 x 14 using partial products.<br>First, break apart the numbers: 23 = 20 + 3, 14 = 10 + 4<br>Now, multiply each part: 20 x 10 = 200, 20 x 4 = 80, 3 x 10 = 30, 3 x 4 = 12<br>Add them up: 200 + 80 + 30 + 12 = 322<br><br>Understanding this will help you with the problem. Want to try the main question now? |

### 5.3 Model selection

We regularly run experiments to determine whether to migrate AI Components of our tutor to different models (OpenAI models, n.d). These include new model releases, different model families, or more compact models. Newer models are not always strict improvements on every dimension important for quality tutoring. Thus, it is important to run live experiments, even if migrating within the same model family. For example, model choice can have a drastic impact on the degree of assistance the AI tutor provides (e.g. Borchers 2025). We initially compare models without any prompt engineering, biasing us in favor of the incumbent model. Follow-up experiments involving model-tailored prompt engineering are conducted as needed.

**Migrating Exercise from GPT-4o to GPT-4.1:** We ran an experiment comparing AI tutor performance when the main completion was generated using GPT-4o vs. GPT-4.1 This was a “do no harm” test, meaning we would launch if the newer model matched the incumbent’s performance.

*Statistically significant changes:*

- Tutor giving away final answer decreased by 96.66% ± 4.66%
- Behavioral engagement increased by 9.93% ± 2.11%

**Migrating Exercise Math Agent from GPT-4o to GPT-4.1:** We ran an experiment comparing AI tutor performance when the Math Agent (see **Fig. 1**) completion was generated using GPT-4o vs. GPT-4.1.

*Statistically significant changes:*

- Next item correctness increased by 3.41% ± 3.41%
- Cognitive engagement increased by 19.29% ± 12.09%
- Behavioral engagement decreased by 3.54% ± 1.80%

**Migrating Tutor Me input classifier from GPT-4.1 to GPT-4.1-mini:** In Tutor Me, students provide a tutoring request to our AI tutor, and we classify the request to determine how to help (see Section 2.2). Following a number of changes to our classification scheme, we wanted to measure classifier accuracy. First, we created an offline evaluation dataset (see Section 3.2). We observed that classification accuracy was higher than expected, so we decided to test switching to a more compact model for lower computational impact and latency.

*Statistically significant changes:*

- Cognitive engagement increased by 11.79% ± 8.52%
- Time to first token latency (P50) decreased by 14.40% ± 1.30%
- Tutoring requests classified as procedural problems with math steps increased 19.10% ± 4.90%, without decreasing the proportion successfully solved by students

# 6 Discussion

In this section, we share some learnings and best practices.

## 6.1 Offline Evaluation vs Live Experiments

The above experiments illustrate the power of live experiments for hill climbing. Currently, we use offline evaluations for smoke tests during development while live experiments are used for hill climbing. In the past, we used to do hill climbing with offline evaluations but that had the following issues:

- Our primary metrics (next-item correctness, cognitive engagement) rely on user actions which cannot be computed in single-turn offline evals.
- It is difficult to build and maintain representative datasets. They need to be stripped of PII and be kept up to date as user behavior drifts.
- In many cases, the datasets saturated very quickly. Our offline metrics would quickly reach > 80% and the confidence intervals for subsequent improvements would be too wide for us to measure any lift. Note that our offline evaluation datasets are of O(100).

Live experiments allow us to get over these limitations. Our current strategy for any hill-climbing change is (1) Conduct offline evals to make sure that the change is ok to show to users. This includes spot-checking threads from the eval datasets (2) Run a live experiment for 1-2 weeks (3) If the experiment shows improvements in metrics, launch the change.

### 6.2 Effect of holidays on Metrics

We saw improvements in our primary metrics during the US holidays of Thanksgiving and Christmas. We believe that this is due to more motivated students using the platform. During the school year, students are obligated to use the platform by their teachers while it is optional during holidays.

### 6.3 Students adapting to changes

Some students on the platform try to cheat by coaxing the AI tutor to give away the answer (as measured by the giving away the answer metric). When we make changes and see improvements in that metric, we find that students quickly find new ways to coax the answer out of the AI tutor. This results in a seesawing giving-away-answer metric, where we eventually lose some (but not all) the gains after the change is put into production.

### 6.4 Need for rapid experimentation

We have often found that our intuition often does not match outcomes. We therefore rely on rapid experimentation to improve AI tutoring quality and would encourage other practitioners in the field to do the same. Above we listed some of the experiments that have moved our metrics significantly, but there were many more: in the 5 months leading up to this paper we conducted over 40 experiments. Most experiments yield small gains but cumulatively the improvements have been significant: next-item correctness by 10%, cognitive engagement by 14%.

**Disclosure of Interests.** The authors have no competing interests to declare that are relevant to the content of this article.